\documentclass[
 preprint,
 amsmath,amssymb,
 aps
]{revtex4-2}

\usepackage{graphicx}
\usepackage{dcolumn}
\usepackage{bm}
\usepackage{multirow} 

\begin{document}

\preprint{APS/123-QED}

\title{Processing-Dependent Near-Field Radiative Heat Transfer at Au/SiC Interfaces}

\author{A. Márquez}
\author{R. Esquivel-Sirvent}%
 \email{raul@fisica.unam.mx}
\affiliation{%
Instituto de Física, Universidad Nacional Autónoma de México, Apartado Postal 20364, México 01000, Mexico\
}%



\date{\today}

\begin{abstract}
Thermal annealing is a widely used thin-film processing technique for modifying interfacial optical losses and electronic scattering in plasmonic materials. Here, we investigate how thermal annealing of gold thin films deposited on silicon carbide substrates influences interfacial near-field radiative heat transfer across nanoscale vacuum gaps. Using experimentally measured dielectric functions for annealed and unannealed Au films, we evaluate the spectral and total radiative heat flux between Au/SiC interfaces within a fluctuational electrodynamics framework.

We show that annealing-induced changes in the low-frequency dielectric losses of Au significantly alter evanescent electromagnetic coupling at the interface, leading to enhancements of up to ~40\% in the total near-field radiative heat transfer at separations of tens of nanometers. Mode-resolved analysis reveals that this enhancement originates from strengthened coupling of overdamped plasmonic surface modes, which are highly sensitive to thin-film processing and interfacial microstructure. These results demonstrate that standard thermal annealing provides a practical route for tuning interfacial radiative heat transfer in metallic thin-film systems without modifying material composition or geometry, offering guidance for the design and interpretation of nanoscale thermal and plasmonic interfaces.
\end{abstract}


\maketitle


\section{Introduction}

Near-field radiative heat transfer between two surfaces refers to the radiative heat exchange when two bodies at different temperatures are separated by a vacuum gap smaller than Wien's thermal wavelength \cite{HARGREAVES}. Unlike the far-field case, described by the Stefan-Boltzmann law for gray bodies, in the near field, radiative heat transfer depends on the separation between the bodies. It can be several orders of magnitude larger than that predicted for black bodies in the far field \cite{Song,Rousseau}. This phenomenon is explained by Rytov's theory of fluctuating electrodynamics \cite{Vinogradov}, which accounts for the contributions of both propagating and evanescent electromagnetic modes between the bodies \cite{pendry1999radiative}.  

The NFRHT has been extensively verified through several experiments  \cite{Song,Rousseau,Ottens,Francoeur15,Gelais}. The interest in understanding the role of different materials, such as metasurfaces,  2D \cite{esquivel2023near,gusso2024near}, and topological materials \cite{didari2024topological},  is due to the potential applications of NFRHT.  Examples include the search for high-power density and high-efficiency energy conversion platforms \cite{mittapally2021near},  controlled heat extraction and energy recycling\cite{bhatt2020integrated},  nanoscale thermal transistors \cite{ben2014near}, thermal logic devices\cite{ben2015heat}, and thermal management \cite{lim2024nanoscale} among others.  These examples are not an exhaustive review, but rather a sampler of recent achievements in near-field heat transfer.

 Two experimental studies on NFRHT  caused some controversy owing to the disparity in their results. The experiments conducted by Reddy et al.\cite{cui2017study} demonstrated good agreement with theory. In contrast, Kittel et al. \cite{kittel} predicted NFRHT values four orders of magnitude above theoretical predictions. Both experiments measured NFRHT for separations less than 10 nm between Au-coated tips and an Au substrate. The differences in results have been attributed to possible sample contamination or other effects, such as phonon tunneling, which are not considered in the current NFRHT theory.  The discrepancies between these experiments demonstrate the need to identify confounders in NFRHT measurements.  Some of  the possible confounders reported in the literature \cite{Song,kim2015radiative} are distance measurement and control \cite{rousseau2009radiative},  residual far field radiation, variations in stability and separation due to the presence of Casimir forces \cite{volokitin2007near}, variations in dielectric properties and film thickness \cite{francoeur2008near}.

 Annealing and quenching of thin metallic films is a common  post-deposition thermal treatment in surfaces used to modify the microstructure and optical properties of the surface \cite{totten1993handbook}. In annealing, the film is heated to an elevated temperature and held for a controlled time, allowing atomic rearrangement that promotes grain growth, reduces internal strain, and decreases defect density. Quenching corresponds to the rapid cooling that follows, which can preserve the modified microstructure or, depending on the cooling rate, introduce new strain and small-scale disorder. 
 
  The annealing and quenching of metallic layers deposited on a substrate change the surface morphology and thus its optical properties depending on such factors  as  annealing time \cite{vsvorvcik2011annealing}, temperature, sputtering time, and substrate where the films are deposited\cite{syed2018thermal,alonzo2013understanding,
vook1965structure}.  The structural and electronic properties of  surfaces \cite{Huang2010} are also affected. This in turn affects their dielectric function. Surface quality and annealing can affect the surface plasmon resonances \cite{sambles1991optical}. This can be explain due to an atomic rearrangement of the surface atoms, promoting grain growth, and the reduction of defects. Quenching,  can lock in these structural changes or, depending on the rate, introduce new strain and defects. 

Most prior studies of near-field radiative heat transfer (NFRHT) have emphasized the role of material selection—such as metals, polar dielectrics, and low-dimensional systems—or geometric design strategies including nanogaps, gratings, and multilayer architectures. While the influence of surface morphology and electronic scattering on optical response is well known  \cite{Song,Rousseau,Ottens,Francoeur15,Gelais}, the specific impact of standard post-deposition thermal processing on near-field radiative exchange has received comparatively less systematic attention. In particular, annealing and quenching are routinely employed in thin-film fabrication to modify microstructure and low-frequency dielectric losses, yet their consequences for evanescent electromagnetic coupling remain incompletely quantified. By directly incorporating experimentally measured dielectric functions of annealed and unannealed Au films into NFRHT calculations, this work clarifies how thermal post-processing influences near-field energy transfer and highlights processing history as a relevant parameter when interpreting and comparing near-field heat-transfer experiments.

\section{Processing-Dependent Dielectric Function of Au Thin Films} 

All dielectric functions used in this work are taken directly from independent experimental measurements of annealed and unannealed Au thin films; consequently, no adjustable or fitted parameters are introduced in the near-field radiative heat-transfer calculations. Specifically, we adopt the experimentally measured optical properties reported by Shen et al. \cite{shen2016temperature}, who systematically characterized the influence of thermal annealing and quenching on the dielectric response of Au over a broad temperature range.

In Ref. \cite{shen2016temperature}, Au films were deposited by electron-beam evaporation using gold with 99.99\% purity. The films were subsequently annealed by heating the samples at 600 K for 15 min. Dielectric function measurements were reported at three temperatures: 300, 450, and 570 K, from which the corresponding optical parameters were extracted and are used here as fixed inputs.

For use in the near-field radiative heat-transfer calculations, the experimentally measured dielectric functions reported in \cite{shen2016temperature} are represented using a Drude–Lorentz model. This is, 
\begin{equation}
\epsilon(\omega)=\epsilon_{\infty}-\frac{f_0\omega_p^2}{\omega^2+i\omega\gamma}+\sum_{m=1}^5\frac{f_m \omega_p^2}{\omega^2_m-\omega^2+i\omega \Gamma_m}, 
\label{dielec}
\end{equation}
which includes the oscillator strength $f_0$,  the high-frequency dielectric function $\epsilon_{\infty}$,  the plasma frequency $\omega_p$, and  the phenomenological electronic damping $\gamma$.  The interband term of the dielectric function consists of the sum of 5 Lorentz resonances \cite{ch1971,rakic1998}.  For  each term $m$ we have the oscillator strength $f_m$, the resonance frequency $\omega_m$ and damping $\Gamma_m$. 

This parametrization provides a compact and frequency-continuous description of the measured optical response and does not introduce additional fitting beyond the experimental analysis itself. The corresponding Drude–Lorentz parameters for annealed and unannealed Au films at the three measured temperatures are listed in Table 1 and are used throughout this work as fixed inputs.
  
\begin{table}[htbp]
\centering
\begin{tabular}{|c|c|c|c|c|c|c|}
\hline
\multirow{2}{*}{\textbf{Temperature (K)}} & \multicolumn{3}{c|}{\textbf{Unannealed}} & \multicolumn{3}{c|}{\textbf{Annealed}} \\ \cline{2-7} 
 & \textbf{300} & \textbf{450} & \textbf{570} & \textbf{300} & \textbf{450} & \textbf{570} \\ \hline
$\varepsilon_{\infty}$ & 1.00 & 1.15 & 1.35 & 1.03 & 1.24 & 1.3 \\ \hline
$\omega_p$ (eV) & 8.22 & 8.63 & 8.023 & 8.12 & 8.072 & 8.023 \\ \hline
$\Gamma_D$ (eV) & 0.046 & 0.057 & 0.113 & 0.086 & 0.099 & 0.117 \\ \hline
$\omega_1$ (eV) & \multicolumn{3}{c|}{4.3} & \multicolumn{3}{c|}{4.3} \\ \hline
$f_1$ & 0.29 & 0.29 & 0.293 & 0.33 & 0.36 & 0.38 \\ \hline
$\Gamma_1$ (eV) & 0.8 & 0.805 & 0.81 & 1 & 1.05 & 1.1 \\ \hline
$\omega_2$ (eV) & \multicolumn{3}{c|}{3.5} & \multicolumn{3}{c|}{3.5} \\ \hline
$f_2$ & 0.0244 & 0.0242 & 0.0244 & 0.019 & 0.0199 & 0.0207 \\ \hline
$\Gamma_2$ (eV) & 0.1 & 0.105 & 0.11 & 0.1 & 0.105 & 0.11 \\ \hline
$\omega_3$ (eV) & \multicolumn{3}{c|}{2.97} & \multicolumn{3}{c|}{2.97} \\ \hline
$f_3$ & 0.151 & 0.148 & 0.146 & 0.128 & 0.13 & 0.132 \\ \hline
$\Gamma_3$ (eV) & 0.6 & 0.605 & 0.61 & 0.6 & 0.605 & 0.61 \\ \hline
$\omega_4$ (eV) & \multicolumn{3}{c|}{2.3} & \multicolumn{3}{c|}{2.3} \\ \hline
$f_4$ & 0.0035 & 0.0095 & 0.015 & 0.0048 & 0.0072 & 0.012 \\ \hline
$\Gamma_4$ (eV) & 0.5 & 0.75 & 0.8 & 0.5 & 0.6 & 0.7 \\ \hline
$\omega_5$ (eV) & \multicolumn{3}{c|}{1.7} & \multicolumn{3}{c|}{1.7} \\ \hline
$f_5$ & 0.0002 & 0.0011 & 0.0024 & 0.0012 & 0.0018 & 0.00305 \\ \hline
$\Gamma_5$ (eV) & 0.4 & 0.45 & 0.5 & 0.4 & 0.45 & 0.5 \\ \hline
\end{tabular}
\caption{Drude-Lorentz parameters for unannealed and annealed gold obtained by Shen \cite{shen2016temperature} .}
\end{table}

\begin{figure}[h]
\includegraphics[width=120mm]{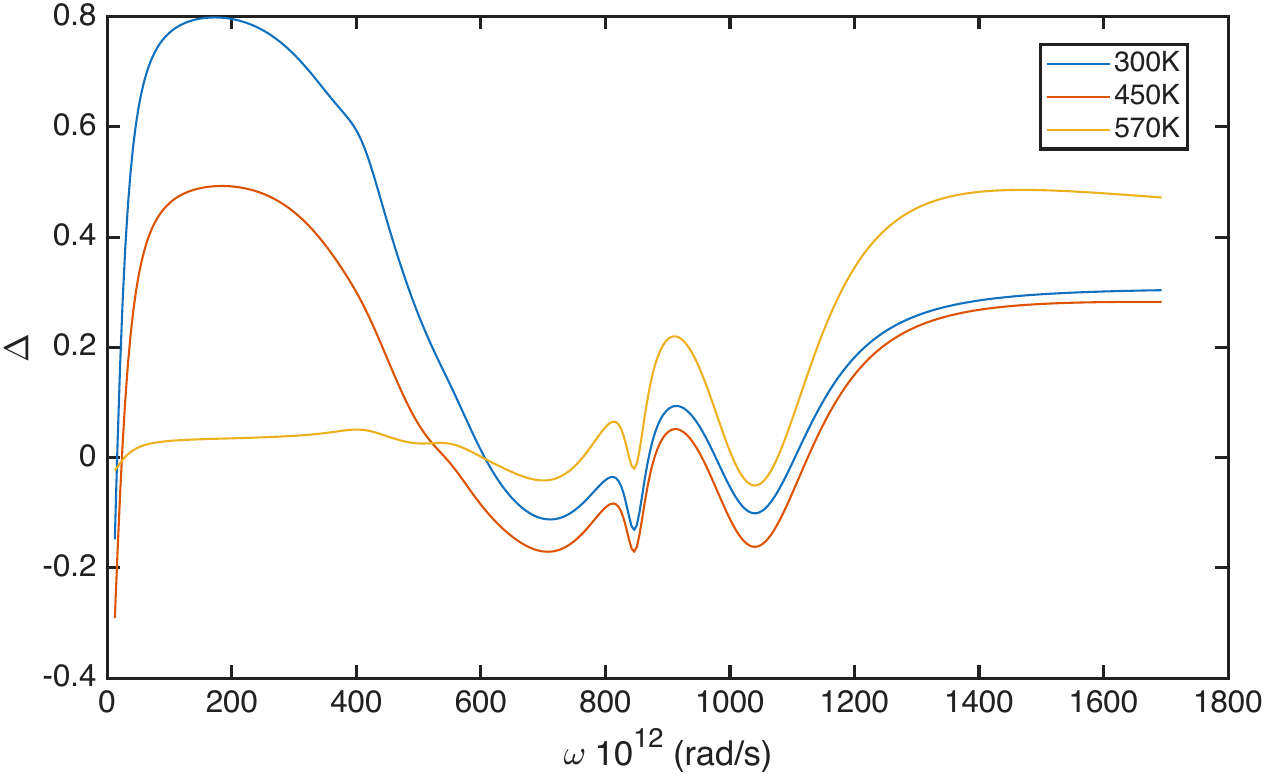} 
\caption{Percent difference between the imaginary part of the dielectric function for annealed and unannealed samples at different temperatures.}
\label{diff}
\end{figure}

To visualize the difference between the annealed and unannealed dielectric functions of the Au films, we  calculate their  difference at a given temperature. This is $\Delta=\frac{\epsilon_{an}^"-\epsilon_{un}^"}{\epsilon_{un}^"}$, where the subscripts refer to the annealed (an) and unanneled (un) dielectric function, and $\epsilon^"$ is the imaginary part. This is presented in Fig.(\ref{diff}). A positive value means The differences in the imaginary part of $\epsilon(\omega)$ are most pronounced in the infrared region, between annealed and unannealed films. 

Although annealing promotes grain growth and reduces certain defects, the experimentally measured dielectric response reflects the net effect of multiple scattering mechanisms, including surface morphology and temperature-dependent electron–phonon interactions, which can lead to increased effective damping in the infrared \cite{yakubovsky2017optical}.

 Another important aspect, that will become relevant in the NFRHT calculation, is that ellipsometry on rough surfaces inherently homogenizes the optical response and  models the roughness as an effective thin layer \cite{bian2023spectroscopic}. This allows fitting to experimental data \cite{KONG1998775}.

\section{Near-Field Radiative Heat Transfer Formalism} 

 We consider two parallel plates, each at a different temperature $T_1$ and $T_2$, separated by a distance $L$. Each plate consists of a SiC substrate and an Au film of thickness $d$ deposited on its surface.  
 
 Near-field heat transfer is calculated using the well-known formalism based on Rytov's theory of thermally excited electromagnetic fields \cite{rytov1959theory,rytov1989principles,pendry1999radiative}, which takes into account the contribution of both propagating and evanescent waves between the plates.  Let $\omega$ be the angular frequency, and define $k$ as the wavevector component perpendicular to the surface in vacuum, given by
$k = \sqrt{\omega^2/c^2 - \kappa^2}$,
where $c$ is the speed of light in vacuum, and $\kappa$ is the wavevector component parallel to the surface. In the Au layer, the perpendicular wavevector component is $k_{\text{Au}} = \sqrt{\epsilon_{\text{Au}} \omega^2/c^2 - \kappa^2},$
and in the SiC layer, it is $
k_{\text{SiC}} = \sqrt{\epsilon_{\text{SiC}} \omega^2/c^2 - \kappa^2},
$ where $\epsilon_{\text{Au}}$ and $\epsilon_{\text{SiC}}$ are the relative permittivities of Au and SiC, respectively.

Let $r_{p,s}^{i}$ be the reflectivity of plate $i = 1, 2$ for $p$- or $s$-polarized waves. The energy transmission coefficients, which describe the coupling between the two slabs for both propagating and evanescent waves, are given by  \cite{rousseau2009radiative,Song,kim2015radiative}              

\begin{align}
\tau_{j=p,s}^{\mbox{prop}}\left(\omega,\kappa,L\right)=\frac{\left(1-|r_{j}^1|^2\right)\left(1-|r_{j}^2|^2\right)}{|1-r_{j}^1r_{j}^2\exp\left(2i k L\right)|^2},\\
\tau_{j=p,s}^{\mbox{evan}}\left(\omega,\kappa,L\right)=\frac{4\mbox{Im}(r_{j}^1)\mbox{Im}(r_{j}^2)\exp\left(-2|k|L\right)}{|1-r_{j}^1r_{j}^2\exp\left(2i k L\right)|^2}.
\end{align}

The total heat flux can be written 
\begin{equation}\label{eq:QT}
Q_T(L,T_1,T_2)=\int_{0}^{\infty}d \omega~ S_{\omega}(\omega, L,T_1,T_2),
\end{equation}
where $S_w$ describes the spectral heat flux following the expression, 
\begin{align}\label{eq:Sw}
S_{\omega}(\omega,L,T_1,T_2)=\left[\Theta\left(\omega,T_1\right)-\Theta\left(\omega,T_2\right)\right]\\ \nonumber
\times\sum_{j=p,s}\int\frac{d\kappa \kappa}{(2\pi)^2}\left[\tau_j^{\mbox{prop}}\left(\omega,k,L\right)+\tau_j^{\mbox{evan}}\left(\omega,k,L\right)\right], 
\end{align}
where $\Theta\left(\omega,T\right)=\hbar \omega/(exp(\hbar \omega/\kappa_B T)-1)$ is the Planckian energy  distribution function at a temperature $T$.  Here, $\hbar$ is Planck's constant and $\kappa_B$ Boltzmanns's
constants. 
For the present work, is useful to define the integrand of Eq.(\ref{eq:Sw}) as the net flux-density. This is 
\begin{align}\label{eq:Hw}
H(\omega,L,T_1,T_2)=\left[\Theta\left(\omega,T_1\right)-\Theta\left(\omega,T_2\right)\right]\\ \nonumber
\times\sum_{j=p,s}\frac{ \kappa}{(2\pi)^2}\left[\tau_j^{\mbox{prop}}\left(\omega,k,L\right)+\tau_j^{\mbox{evan}}\left(\omega,k,L\right)\right].
\end{align}
The term $H(\omega,L,T_1,T_2)$ is indicative of how much energy is carried per electromagnetic mode of frequency and wavevector, and is a convenient mode-resolved diagnostic. 

\section{Results}
In the calculations, we considered the temperatures of each plate to be $T_1=450$ K and $T_2=300$ K. 
Since Au films are deposited on a SiC substrate, the reflection coefficients $r_{p,s}$ are calculated using the expressions \cite{yeh2006optical}, 

\begin{equation}
r_{p,s}^i=\frac{r_{p,s}^{01}+r_{p,s}^{12}e^{2ik_{Au}d}}{1+r_{p,s}^{01}r_{p,s}^{12}e^{2ik_{Au}d}},
\end{equation}
where $r_{p,s}^{01}$ and $r_{p,s}^{12}$ are the Fresnel coefficients between vacuum and Au and between Au and SiC, respectively. It is essential to note that the dielectric function varies with temperature, which in turn alters the reflection coefficients for the plates at different temperatures. Also, we consider a 110 nm  Au layer deposited on a SiC substrate. For thicknesses equal to, or larger than,  110 nm, the dielectric function of Au can be described by its bulk value \cite{esquivel2016ultra, walther2007terahertz}. Otherwise, corrections to the plasma frequency have to be considered. 

\begin{figure}[h]
\includegraphics[width=170mm]{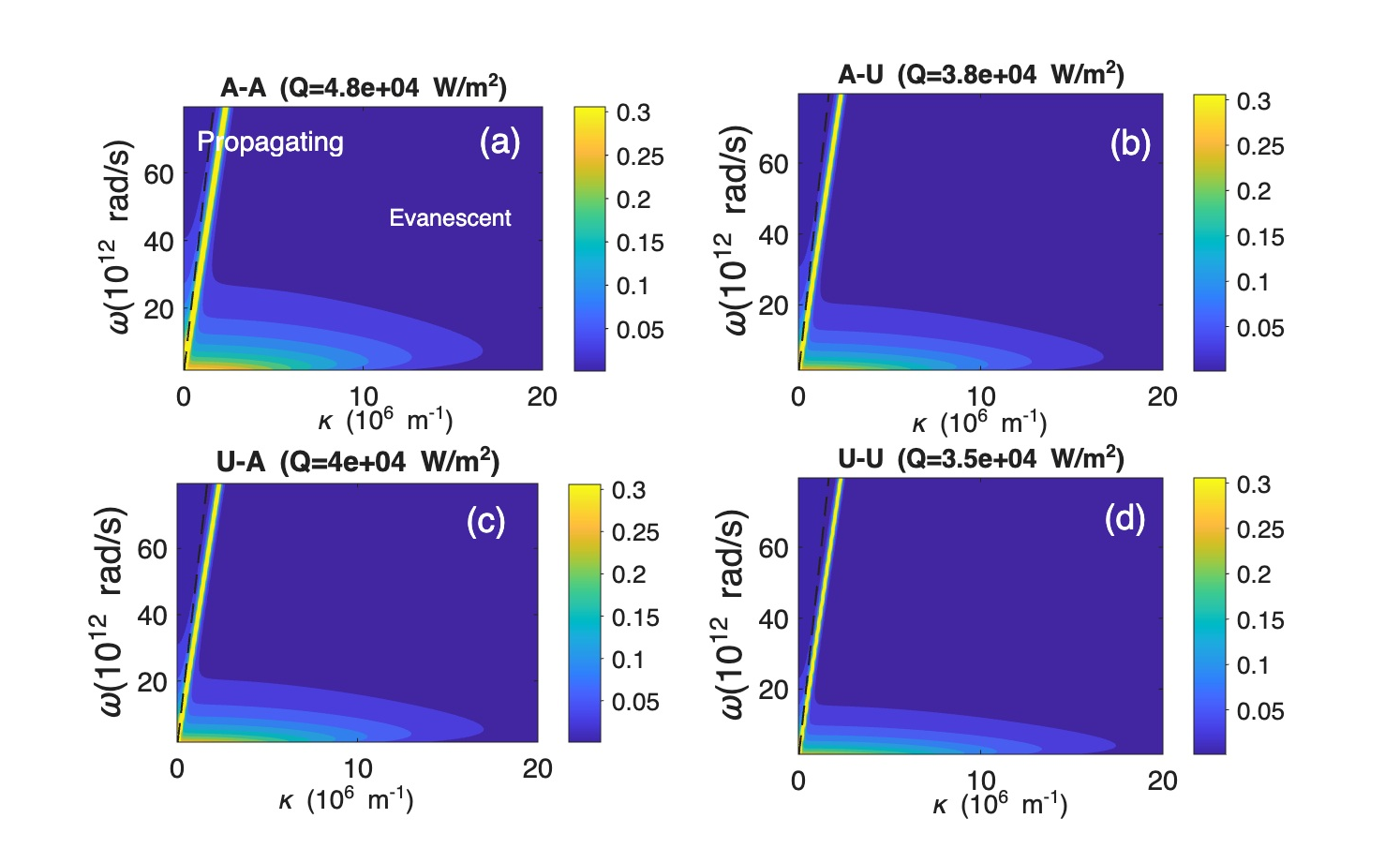} 
\caption{ Energy transmission coefficients (color bars) for near-field radiative heat transfer between $110$nm thick Au films on SiC. The temperature of the hot plate is $T_H=450$ K and $T_C=300$ K (a) unannealed and (b) annealed cases, and (c) their difference. The separation between the plates is L = 50 nm. }
\label{taus}
\end{figure}

First, in Fig.(\ref{taus}), we present the value of the energy transmission coefficients in both the propagating and evanescent case, Eq (2) and Eq (3) respectively, as these are the main contribution to the NFRHT. We consider four different cases: (a) Both surfaces annealed (A-A); (b) Annealed at $T_1$ and unannealed at $T_2$ (A-U); (c) unannealed at $T_1$ and annealed at $T_2$ (U-A); (d) both surfaces unannealed (U-U). The calculations were performed at a separation distance of $50$ nm and the total heat flux, calculated using Eq. (\ref{eq:QT}),  is shown as reference in each of the panels. 

In Fig.(\ref{taus}), the light cone is presented as the black dashed line, separating propagating and evanescent modes. Although the behavior of the energy transmission coefficient is similar, comparing all cases, the width and frequency of the evanescent modes is different. The effect of this difference becomes evident when we compare the total heat flux $Q$ for the A-A case with to the U-U showing a $40\%$ difference. The total heat flux value is shown on top of each panel. In all cases the differences come at low frequencies, below the transverse frequency of SiC. This means that most of the contributions to the NFRHT comes from overdamped plasmon modes in the Au/SiC systems. This is consistent with the results of thin metallic films \cite{chapuis2008effects}.

\begin{figure}[h]
\includegraphics[width=160mm]{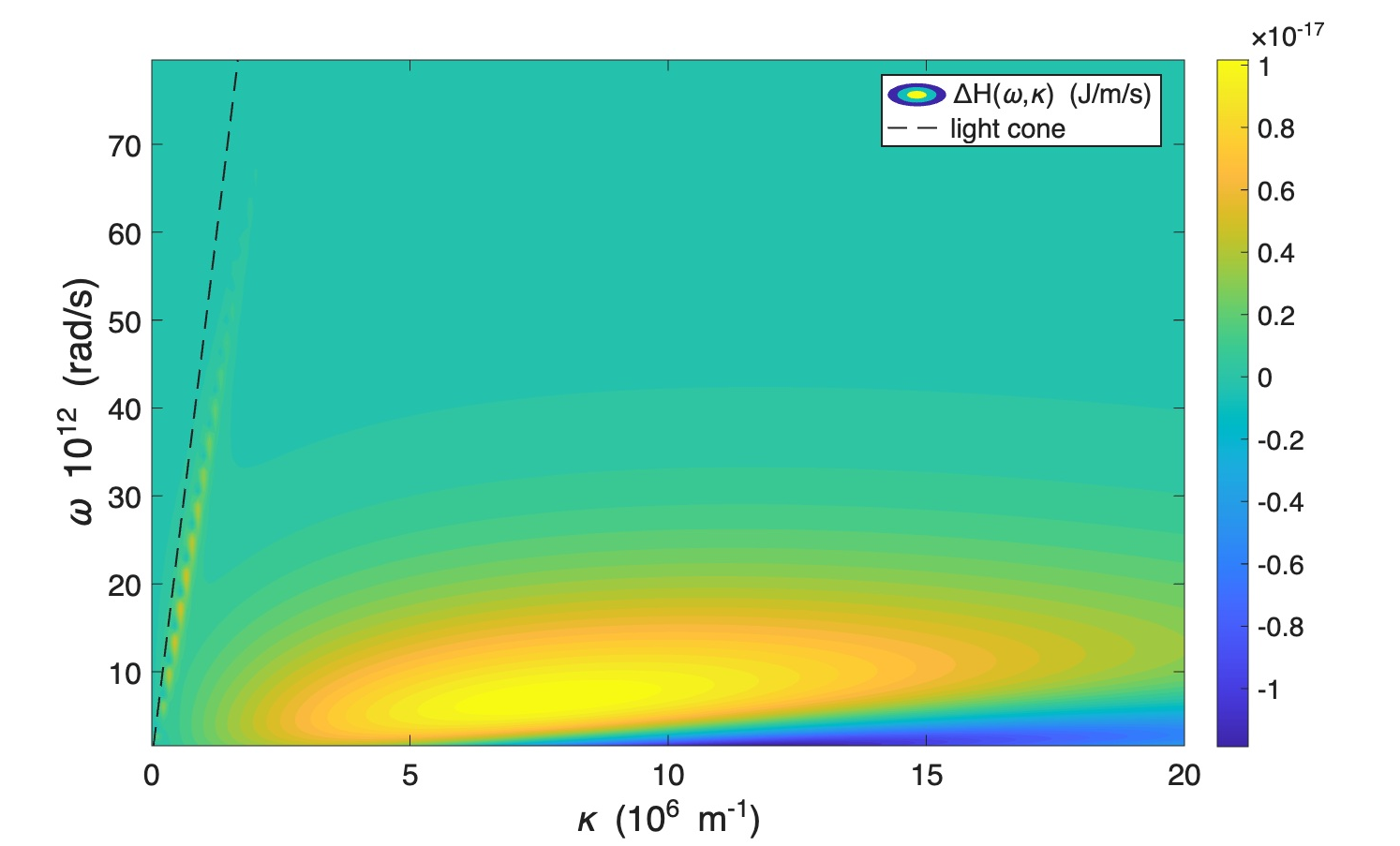} 
\caption{Difference $\Delta H=H_{A-A}-H_{U-U}$  indicating that annealing strengthens evanescent, plasmon-mediated heat transfer (Yellow regions). The negative regions (blue) correspond to suppressed channels. The values of $\Delta H$ are given by the color bar and has dimensions of $W/m^2$}
\label{dh}
\end{figure}

To see more clearly the effect of temperature treatment, using Eq. (6),  the difference on the mode density $\Delta H=H_{A-A}-H_{U-U}$ is calculated and shown in Fig.(\ref{dh}). Figure 3 shows that, while in a wide range the difference between annealed and unannealed materials is zero or close to, there are two regions where the behavior differs. In the yellow region, the positive valued region, heat transfer increases due to the annealing, while in the blue region, the negative valued region, annealing diminishes the heat transfer. These two regions represent competing mechanisms, which once integrated  over frequency and wave-vector result in a higher total heat transfer for the annealed case.

\begin{figure}[h]
\includegraphics[width=110mm]{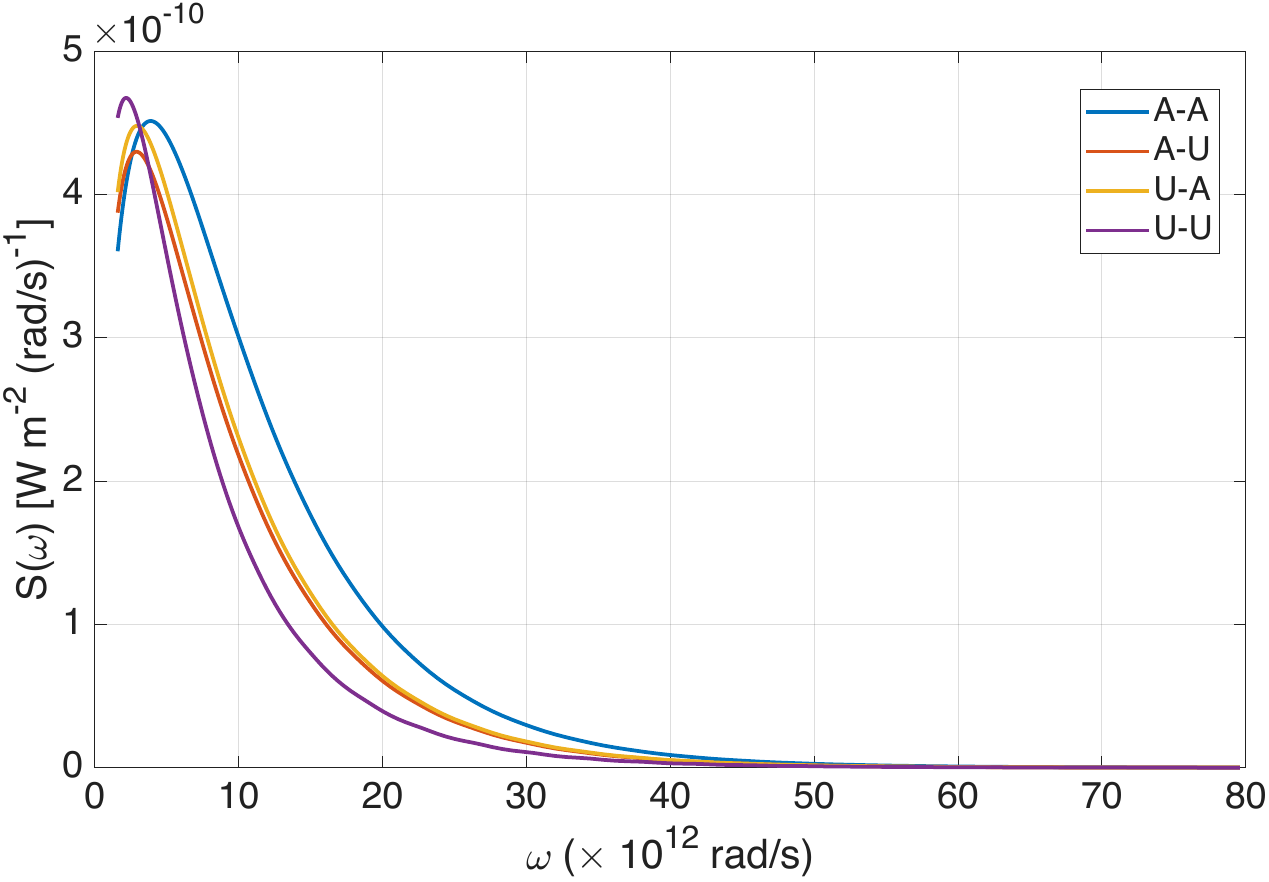} 
\caption{Spectral heat flux $S(\omega)$ for two Au layers separated by $L = 50\,\mathrm{nm}$ at $T_H = 450\,\mathrm{K}$ and $T_C = 300\,\mathrm{K}$. Curves correspond to annealed-annealed (A-A), annealed-unannealed (A-U), unannealed-annealed (U-A), and unannealed-unannealed (U-U) interfaces. Annealing enhances the overall spectral magnitude and shifts the integrated heat flux accordingly.} 
\label{Spectral}
\end{figure}

The spectral heat flux $S(\omega)$ is presented in  Fig.~\ref{Spectral} for a vacuum gap of $L = 50\,\mathrm{nm}$ between two Au layers under the four interface conditions considered previously: A-A, A-U, U-A, and U-U, with $T_H = 450\,\mathrm{K}$ and $T_C = 300\,\mathrm{K}$. The spectra are dominated by low-frequency contributions, displaying a broad maximum in the low-THz range followed by a rapid decay above $\sim 20\,\mathrm{THz}$. This decay is consistent with the Planckian distribution that governs thermal emission in fluctuational electrodynamics, which suppresses high-frequency contributions to the near-field heat transfer. Annealing increases the spectral magnitude across the entire frequency range, with the overall ordering $S_{\mathrm{A\!-\!A}} > S_{\mathrm{A\!-\!U}} \approx S_{\mathrm{U\!-\!A}} > S_{\mathrm{U\!-\!U}}$. The asymmetry on A-U compared with U-A corresponds to the dielectric functions being dependent on the temperature. This ordering agrees with the broader and more intense evanescent transmission observed in Fig.~\ref{taus}, accounting for the $\sim 41\%$ enhancement in the total flux for the A-A configuration relative to U-U.

\begin{figure}[h]
\includegraphics[width=100mm]{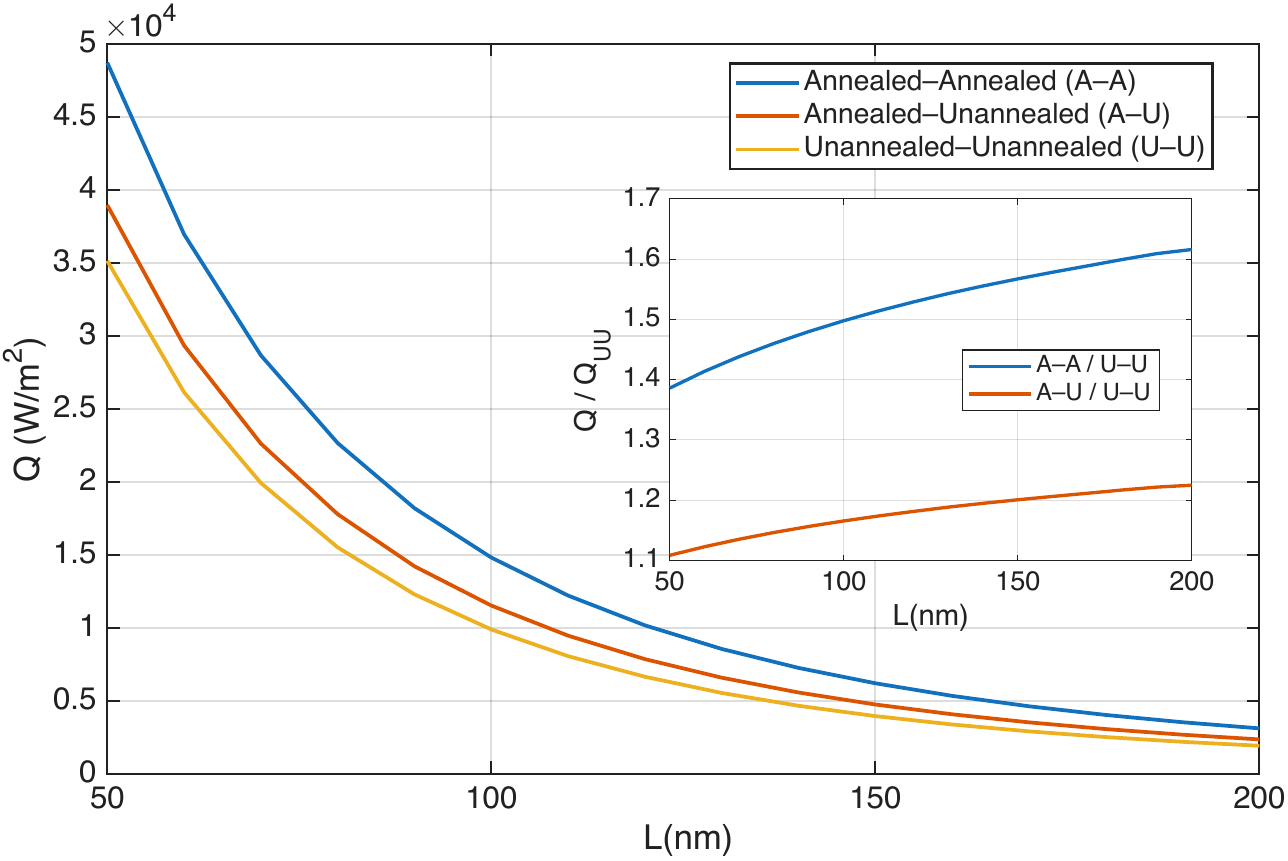} 
\caption{ Total heat transfer as a function of the separation. The temperature difference between the Au surfaces is 150 K. Annealing the surfaces results in higher heat transfer. }
\label{Qtotal}
\end{figure}

Furthermore, the total heat transfer between the surfaces is calculated as a function of the separation $L$ and is shown in Fig. (\ref{Qtotal}). The annealed surfaces exhibit a higher heat flux compared to the unannealed case. The increase in total heat transfer is consistent with the optical response of Au in the infrared range, where the imaginary part of the dielectric function is actually higher for the annealed case. This suggests that, in the relevant near-field regime dominated by evanescent modes, annealing leads to enhanced electromagnetic losses that increase modal coupling and radiative heat exchange. This behavior arises from changes in electron scattering and the electronic density of states at the surface as a result of thermal treatment, which modify the dielectric response of Au.   For clarity, and to better quantify the enhancement after thermal treatment of the Au surface, the inset in Fig. (\ref{Qtotal}) shows the ratios  $Q_{AA}/Q_{UU}$ and $Q_{AU}/Q_{UU}$.

Another critical factor is the change in surface morphology. Annealing of the surface causes the gold grains to aggregate, resulting in surface morphology variations with characteristic length scales \cite{shen2016temperature}, which can lead to possible diffraction effects that are not accounted for. This also explains why, at large separations, both results are similar since $L$ will be larger than any mean changes in roughness.

\section{Conclusions}
In this work, we have investigated how thermal annealing of gold thin films modifies near-field radiative heat transfer across nanoscale vacuum gaps through changes in their electromagnetic response. Using independently measured dielectric functions for annealed and unannealed Au films deposited on SiC substrates, reported previously in the literature, we computed the spectral and total radiative heat flux within the framework of fluctuational electrodynamics.

We find that annealing-induced modifications of the low-frequency dielectric losses of Au lead to significant changes in evanescent electromagnetic coupling, resulting in enhancements of the near-field radiative heat flux of up to approximately $\sim 40$\% at separations of tens of nanometers. A mode-resolved analysis shows that this enhancement is associated with strengthened contributions from overdamped-plasmonic surface modes, whose spectral weight is particularly sensitive to thin-film processing.

These results demonstrate that standard thermal post-processing, routinely employed in thin-film fabrication, can measurably influence near-field radiative heat transfer without altering material composition or geometry. From an experimental standpoint, our findings imply that variations in annealing history and film quality should be carefully accounted for when comparing near-field heat-transfer measurements performed on nominally identical metallic surfaces.

\begin{acknowledgments}
We thank S. G. Castillo-López for valuable comments and discussions. A.M. acknowledges partial funding from 1104375 SECIHTI scholarship.
\end{acknowledgments}

{\bf Conflict of Interest:} The authors have no existing conflict of interest to report. 

\providecommand{\noopsort}[1]{}\providecommand{\singleletter}[1]{#1}%

\end{document}